\documentclass[twocolumn]{aastex631}

\usepackage{color}
\usepackage{amsmath}
\usepackage{changepage}
\usepackage{float}

\graphicspath{{./}{figures/}}

\begin{document}

\title{Identifying Acoustic Wave Sources on the Sun. II. Improved Filter Techniques for Source Wavefield Seismology}

\correspondingauthor{Shah Mohammad Bahauddin}
\email{shahmohammad.bahauddin@colorado.edu}

\author{Shah Mohammad Bahauddin}
\affiliation{Laboratory for Atmospheric and Space Physics, University of Colorado, Boulder, CO 80303, USA}
\affiliation{DKIST Ambassador, National Solar Observatory, Boulder, CO 80303, USA}

\author{Mark Peter Rast}
\affiliation{Department of Astrophysical and Planetary Sciences, University of Colorado, Boulder, CO 80309, USA}
\affiliation{Laboratory for Atmospheric and Space Physics, University of Colorado, Boulder, CO 80303, USA}

\begin{abstract}

In this paper we refine a previously developed acoustic-source filter~\citep{2021ApJ...915...36B}, improving its reliability and extending its capabilities. We demonstrate how to fine-tune the filter to meet observational constraints and to focus on specific wavefront speeds. This refinement enables discrimination of acoustic-source depths and tracking of local-source wavefronts, thereby facilitating ultra-local helioseismology on very small scales.  By utilizing the photospheric Doppler signal from a subsurface source in a MURaM simulation, we demonstrate that robust ultra-local three-dimensional helioseismic inversions for the granular flows and sound speed to depths of at least 80 km below the photosphere are possible.  The capabilities of the National Science Foundation's new Daniel K. Inouye Solar Telescope (DKIST) will enable such measurements of the real Sun.  

\end{abstract}

\keywords{The Sun (1693) --- Solar physics (1476) --- Solar photosphere (1518) --- Helioseismology (709) --- Solar granulation (1498) --- Solar oscillations (1515)}

\section{Introduction} \label{sec:intro}

Solar acoustic waves are thought to be emitted by discrete dynamical events in and below the solar photosphere.  These waves couple to the global modes, and their high frequency components propagate upward into the solar atmosphere where they may play an important role in heating~\citep[e.g.,][]{1946NW.....33..118B, 1948ApJ...107....1S, 2007ApJ...671.2154K, 2020A&A...642A..52A, 2023ApJ...945..154M}.  Multiple mechanisms for the excitation of solar acoustic waves have been proposed~\citep[e.g.][]{1952RSPSA.211..564L, 1954RSPSA.222....1L, 1967SoPh....2..385S, 1991LNP...388..195S, 1994ApJ...424..466G, 1995ApJ...443..863R, 1999ApJ...524..462R}, with some work pointing to the importance of Reynolds-stress induced pressure fluctuations by nearly sonic or slightly supersonic turbulence and other suggesting an important role for radiative cooling and downflow plume formation in the photosphere. There is observational support for both of these mechanisms~\citep[e.g.,][]{1995ApJ...444L.119R, 1998MNRAS.298L...7C, 1998ApJ...495L..27G, 1999ApJ...516..939S, 2000ApJ...535.1000S, 2000ApJ...535..464S, 2001ApJ...561..444S, 2010ApJ...723L.134B, 2010ApJ...723L.175R, 2013JPhCS.440a2044L}, but it is still unclear which of them is dominant, how well each couples to the global p-modes, and whether one or the other is more important for generating the waves that contribute to atmospheric heating. 

Identification and careful characterization of solar acoustic sources requires separating the source and the local wave field it produces from the background convective flows and p-modes. This is difficult because the amplitudes of the individual sources and the emitted waves are typically much smaller than those of the granulation and the p-mode coherence patches (with signal-to-noise ratios, SNRs, well below unity).  Moreover, the spectral content of the source signal overlaps that of the acoustic modes and, in part, also that of the background granular motions. Consequently, direct observation of the source wavefield remains challenging and linear image filtering and frequency domain noise reduction techniques fail in direct detection.

Despite these difficulties, progress has recently been made~\cite[]{2021ApJ...915...36B} in identifying acoustic source sites in an MPS/University of Chicago Radiative MHD~\citep[MURaM; ][]{2005A&A...429..335V, 2009ApJ...691..640R, 2014ApJ...789..132R} simulation of solar convection, based solely on their photospheric signatures. In that work, a convolutional filter was developed that, via high frequency temporal differencing, suppressed the granulation background and allowed direct visualization of the local source wavefield in the simulated photospheric time series. Several key features of the acoustic sources were revealed:  they are clustered on mesogranular scales, usually occur in intergranular lanes, and often at depths corresponding to $\sim50\%$ hydrogen ionization~\citep{2021ApJ...915...36B}.  With the NSF's Daniel K. Inouye Solar Telescope~\citep[DKIST][]{2020SoPh..295..172R} now being commissioned, the high spatial resolution and temporal cadence required by the filter will be achievable with observations.  This will allow determination of whether these same properties hold for the real Sun.

Although powerful, the acoustic source filter developed in ~\cite{2021ApJ...915...36B} has some problems. First, convolutional filters carry the inherent risk that the result one achieves is biased by the convolution one applies, that the pattern one is looking for is imprinted on the data by the convolution itself, and second, the neural network that underlies the filter was trained on the ideal Green's function response, but the wave front observed is often only approximates that solution. These issues can produce false positive and false negative detections. In this paper, we demonstrate that both can be mitigated by eliminating the convolution component of the filter and applying the temporal-differencing directly to the image time series. This allows us to not only identify the source site, as with the previous filter, but to also directly measure the wave-front amplitude, phase speed, and distortion.  We show that the evolution of the wave-front position, can be determined from the difference-image timeseries and used successfully in a helioseismic inversions for the atmospheric properties at very small scales.  

\section{Wavefront filtering techniques} \label{sec:Improvement}

A Doppler-image timeseries of solar photosphere shows large amplitude fluctuations associated with granulation, structured magnetic fields, and modal oscillations.  When sufficiently well resolved, it also shows, with much lower amplitudes, transient fluctuations due to magneto-hydrodynamic waves and shocks.  We are here concerned with identifying propagating wavefronts associated with spatially compact and temporally discrete acoustic wave sources. These transients can be identified against the high amplitude background fluctuations because they occur on shorter time scales compared to the evolutionary time of the granulation, magnetic fields, or $p$-mode coherence~\citep{2021ApJ...915...36B}.  In other words, the later components comprise of a relatively slowly varying background on top of which the transient propagation occurs. Extraction of the acoustic transient signal from an observed time-series thus requires  observations with: (1) adequate spatial resolution, (2) high temporal cadence, and (3) sufficient local signal to noise. 

For direct detection, a discernible signal can be expected if the perturbation associated with the transient has, at a minimum, a higher amplitude than the temporal variance of the background over the time period of the transient wavefront crossing (local SNR equal to one). Under this circumstance, identification of the acoustic perturbation depends only on the spatial pixel scale and the temporal cadence. If the pixel scale of the observations is taken to be fixed, the temporal increment between frames can be tuned for the optimal detection of the propagating wave-front.

For a typical acoustic source within the MURaM photosphere, the width of a transient impulsively-generated acoustic-wavefront response is approximately 60 - 100 km (with some dependence on both the source physics and depth).  Nyquist sampling of this front thus requires a minimum spatial pixel scale of about ${\Delta x} =$ 30 km. Since the sound speed of the solar photosphere $c_s \sim8$ km/s, a temporal resolution of ${\Delta x}/c_s \simeq $ 4 seconds is needed to resolve the wavefront as it travels. In other words, the minimal temporal sampling interval should be ${\Delta t}/2 \simeq$ 2 seconds.  The spatial pixel scale in the simulations we employ is 16 km and the temporal interval between saved snapshots is 2 seconds, so the impulsive wavefront is resolved in both space and time. The Daniel K. Inouye Solar Telescope (DKIST) will be able to achieve similar values~\citep{2020SoPh..295..172R}.

When the amplitude of the acoustic-source wavefront perturbation is smaller than the variance of the background (SNR less than one) it is necessary to employ filtering techniques that amplify the signal (the acoustic wavefront perturbation) and/or suppress the noise (the $p$-modes and convective background). Since, the local source-generated wavefront is usually orders of magnitudes weaker than the background, and since its frequency content overlaps that of the p-modes, and in part the granulation, methods other than linear filtering and frequency domain noise reduction are needed. 

\subsection{Critical assessment of previously proposed convolutional filter} \label{sec:old}

In our previous work~\citep{2021ApJ...915...36B}, we employed a convolutional neural network to identify sites of local acoustic emission in a MURaM simulation of the solar photosphere.  The neural network was trained to identify propagating wavefronts produced by sources even when the wavefront to background SNR was less than one.  In order to train the neural network, we utilized, as a local source response template, the Green's function solution to the wave equation in two dimensions.  Using this template, a convolutional neural network was trained to provide a probabilistic assessment of whether a source is located at a given place and time based on the idealized propagating wavefront signal that such a site should produce. Following successful training, we unwound the interlaced convolutional kernels of the deep-learning algorithm, and deconstructed them into a set of linearly summed traditional operators. That strategy enabled us to convert the convolutional neural network into an image filter.  It entailed the sequential application of a spatiotemporal convolution followed by temporal convolution with a threshold clipping. Importantly, the temporal convolutional kernel that the neural network defined was equivalent to an $n$-difference filter in time~\citep{2021ApJ...915...36B}.  

While powerful, subsequent work has uncovered some drawbacks to this method. These arise for two fundamental reasons. The convolutional neural network was trained on the full spatio-temporal extent of the expanding wavefront (the Green's function solution).  Wavefronts generated by and in the presence of vigorous granular flows are, however, often spatially incomplete or significantly asymmetric in amplitude; often only a portion of the expanding front is visible.  For example, a wave generated by a source embedded in an intergranular lane often manifests as a partial wavefront propagating across the neighboring granule to one side but not the other. The convolutional kernel, defined by a neural network trained on the complete Green's function solution, poorly maps onto these partial wavefronts. Additionally, the expansion of small granular structures can mimic the circular wavefront kernel when expansion velocities are close to the local sound speed. The first of these results in false negative detections while the latter produces false positives.

More generally, as mentioned previously, convolutional filters carry the risk that the pattern one is looking for is imprinted on the data by the convolution itself.  We have found that, after applying our previous method, strong false positive detections occur at sites of rapidly-evolving localized dynamics, which approximate Dirac delta-function-like impulses in the image time series.  If such impulses in the image time-series spuriously arises from an observational artifact, such as a energetic-particle hit on the detector or an abrupt explosive event on the Sun, they can be identified by careful analysis and the neural network can be trained to register such events as a false positives. On the other hand, if the apparent impulse arises in the Doppler map due to rapid dynamics, such as a vanishing granule or very sharp downflow amplification in an intergranular lane, it is much more difficult to objectively define and detect, and it is very difficult to train a convolutional neural network to avoid it. Such dynamical events are thought to be possible source of acoustic emission, so it is particularly important not to bias the wavefront detection in their favor simply because they are spatially and temporally localized.

\subsection{Improved filtering technique} \label{sec:new}

Fortunately, these difficulties can be mitigated by eliminating the spatiotemporal convolutional component of the previously developed filter and applying the neural-network motivated temporal-difference directly to the image time series.  Moreover, direct application of the difference filter to the image time series allows it to be tuned to the propagating wavefront speed, facilitating selection of those wavefronts with specific phase speeds. Since the phase speed of the wavefront depends on the source depth, the atmospheric sound-speed stratification, and the flows through which the wavefront is propagating, measuring the propagation and distortion of a wavefront of a particular phase speed allows inversion for the atmospheric properties through which that wavefront has propagated (see~\S\ref{sec:SWS} below).  

In this section, we will demonstrate why temporal difference filters are useful for extracting low-SNR acoustic waves propagating through and in the solar photosphere. We begin by highlighting that the application of an $n$-difference filter to an image timeseries is equivalent to taking successive derivatives of the signal using a forward finite-difference scheme of first-order accuracy.  Such an operation amplifies high temporal-frequency components of the images and suppresses slowly evolving ones. Since, the $n$-difference filter measures the perturbation over an $n-$point stencil, a well-resolved slowly evolving perturbation of nearly constant value across the stencil returns a nearly null value.  A signal evolving quickly, on the other hand, returns a difference measure of large absolute value. Thus, the $n$-difference filter combines two aspects of canonical motion detection algorithms: temporal differencing~\citep[e.g.,][and references therein]{5209870}, which in motion detection applications typically employs only adjacent timesteps~\citep[a first derivative, though cf.,][]{4766907, Paul2017}, and background subtraction, implicitly included in our scheme by the higher derivative suppression of non-evolving or slowly-evolving image contributions.  Importantly, direct application of the temporal difference filter to the image time series, outside of the convolutional neural network, removes the risk of spurious source introduction by the spatiotemporal convolution discussed above. 

In the solar photosphere, there are two primary contributions to Doppler images that act as background noise: granular motions and the modal oscillations. Largely obscured by these are the propagating acoustic wave fronts that originate with discrete sources.  Fourier spectra of Doppler-image time series show  power distributed in two major lobes: a lower frequency lobe, lying below the photospheric sound speed, due to granulation, and a higher frequency lobe, composed of ridges, or in short time-series unresolved ridges, due to modal oscillations.  This is the basis for the well-know subsonic filter~\citep[e.g.,][]{1988ApJ...333..996H, 1989ApJ...336..475T, 1998ApJ...505..390S}, that is used to separate granular motions from $p$-modes. Localized acoustic sources produce impulsive responses that propagate horizontally in the photosphere with amplitudes and phase speeds that depend on the source depth (Figure~~\ref{fig:figure4}).  For source depths ranging from about 1.6 Mm below the photosphere to within the photosphere itself, the wavefronts in the photosphere propagate with speeds ranging from about 8 to 13 km/s (mean sound speed on the MURaM $\tau=1$ surface is 7.75 km/s).  Since the amplitude of a source-generated wavefront in the photosphere decreases rapidly with source depth, a filter that removes all power above a line with slope $13$ km/s removes the bulk of the solar $p$-mode power while leaving the granulation and most of the potentially detectable source-generated wavefronts largely intact. The wavefronts produced by sources deeper than 1.6Mm, and thus with phase speeds higher than $13$ km/s, have extremely small amplitudes in the photosphere due their expansion with distance from the source location, and are thus largely undetectable there.  Hence, unless the amplitudes of deep sources in the Sun is very much higher than they are in MURaM, the $13$ km/s cut-off has little effect on source detection, and one can remove most of the $p$-mode contribution with a $13$ km/s low-pass Fourier filter. The remaining Doppler signal includes both granular motions and the local source wavefield, leaving the challenge of removing the granulation contribution.

\begin{figure}[t!]
\centerline{\includegraphics[width=\linewidth]{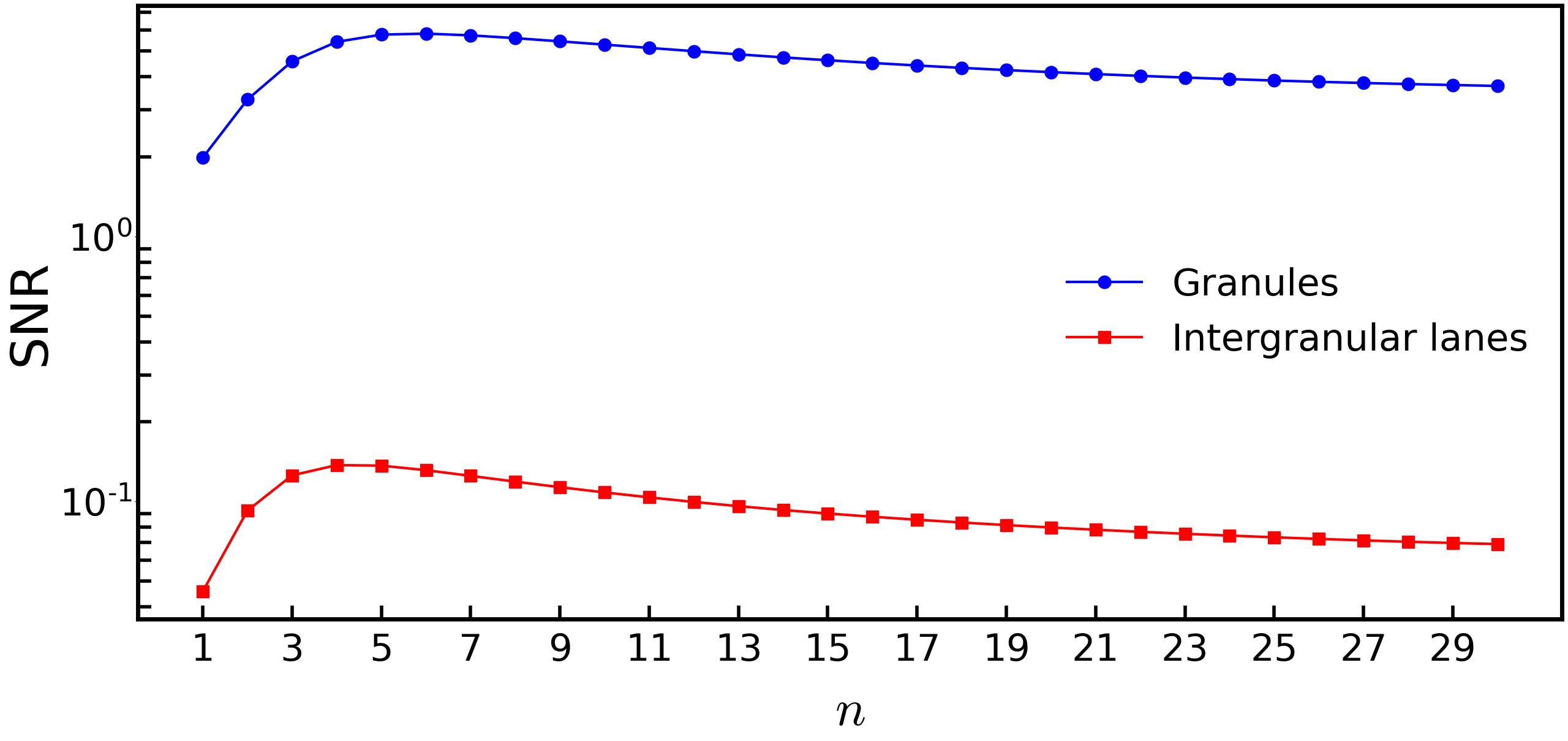}}
\caption{Ratio of the amplitude of a propagating wave to the standard deviation of the granular fluctuations (SNR) after application of an $n$-difference filter, as a function of $n$. In the upflow regions of granulaiton, the SNR quickly rises for $n \le 3$ and plateaus near its maximum $n \ge 5 - 6$.  These are similar values to those deduced from a convolutional neural network~\citep{2021ApJ...915...36B}.}  
\label{fig:figure1}
\end{figure}

Here we apply the $13$ km/s low-pass Fourier filter to better illustrate the effectiveness of the $n$-difference filter in eliminating granulation noise, but note that this Fourier filtering is not always necessary and may be undesirable.  We discuss its benefits and drawbacks further in Section~\ref{sec:summary} below. After applying the $13$ km/s filter, we can compare the amplitude of an idealized wave signal to the standard deviation of the granular motions when both are measured after application of the $n$-difference filter.  In detail, the wave signal is taken to be vertical velocity of the two-dimensional Green's function solution for the wave-front emitted from a photospheric source after it has traveled 500 km horizontally at the mean photospheric sound speed. The source is taken to be Gaussian with a spatial $\sigma = 16$ km and a temporal $\sigma= 2$ s, and an amplitude of 1 km/s.  With this initial condition, the wave-front amplitude of the Green's function solution 500 km from the source is approximately equal to that of the propagating-wavefronts observed in the simulations~\citep{2021ApJ...915...36B}.  The SNR plotted in Figure~\ref{fig:figure1} is the ration of the standard deviation of the $n$-difference of this wave signal to the standard deviation of the granulation in the $13$ km/s Fourier filtered MURaM timeseries, with the standard deviation of the MURaM time series computed separately for the granules and intergranular lanes.

It is evident from Figure~\ref{fig:figure1} that optimal-differencing order $n$ (highest SNR) differs for wave propagating across granules than for those propagating across intergranular lanes. Wavefronts are much more readily detected against the smoother and more slowly evolving background of a granule than against the more complex and rapidly evolving flows in the intergranular lanes, with peak SNR after differencing equal to $6$ in granular regions and $0.14$ when the wave is viewed against the intergranular lane background. With a sampling period of $2$ second, the fastest sampling rate available for the MURaM data cube we analyze, the half-power point ($0.707$ relative to peak) of the SNR over granules occurs when an $n=3$ difference is applied. Although it is possible to obtain better SNR by using higher $n$, high $n$-differences can introduce artifacts, and it is advantageous to keep the $n$ as low as possible while still gaining significant reduction of the granulation contribution.  For the remainder of this paper, we focus on the $3$-difference filter and its use in measuring wavefront propagation across regions of granular upflow.

\subsection{Filter fine tuning } \label{sec:next}

As previously indicated, the difference filter we propose can be tuned to account for the the temporal cadence of the observations, the evolutionary time scale of the background flow, and the phase speed of the small-amplitude source-generated wavefronts in the photosphere.  A successful $n^{th}$ derivative forward-difference scheme measures the change in the signal at each spatial location using $n+1$ time steps over a time period during which the background is nearly unchanged and the signal changes significantly. An over-sampled signal has nearly constant value over the stencil and returns very low amplitude differences.  The same is true of  a well-resolved signal, which changes only a small amount over the $n$ timesteps that contribute to the difference, and so also returns a low value under the $n$-difference operation. A signal with evolves fast compared to the sampling period, on the other hand, returns a large value. However, because the $n$-difference operation measures the signal using an $n+1$ point stencil, it also yields a small-amplitude result if the signal is under-sampled, in which case the difference signal is under resolved and has non-zero values over only a fraction of the stencil. 

\begin{figure}[t!]
\centerline{\includegraphics[width=\linewidth]{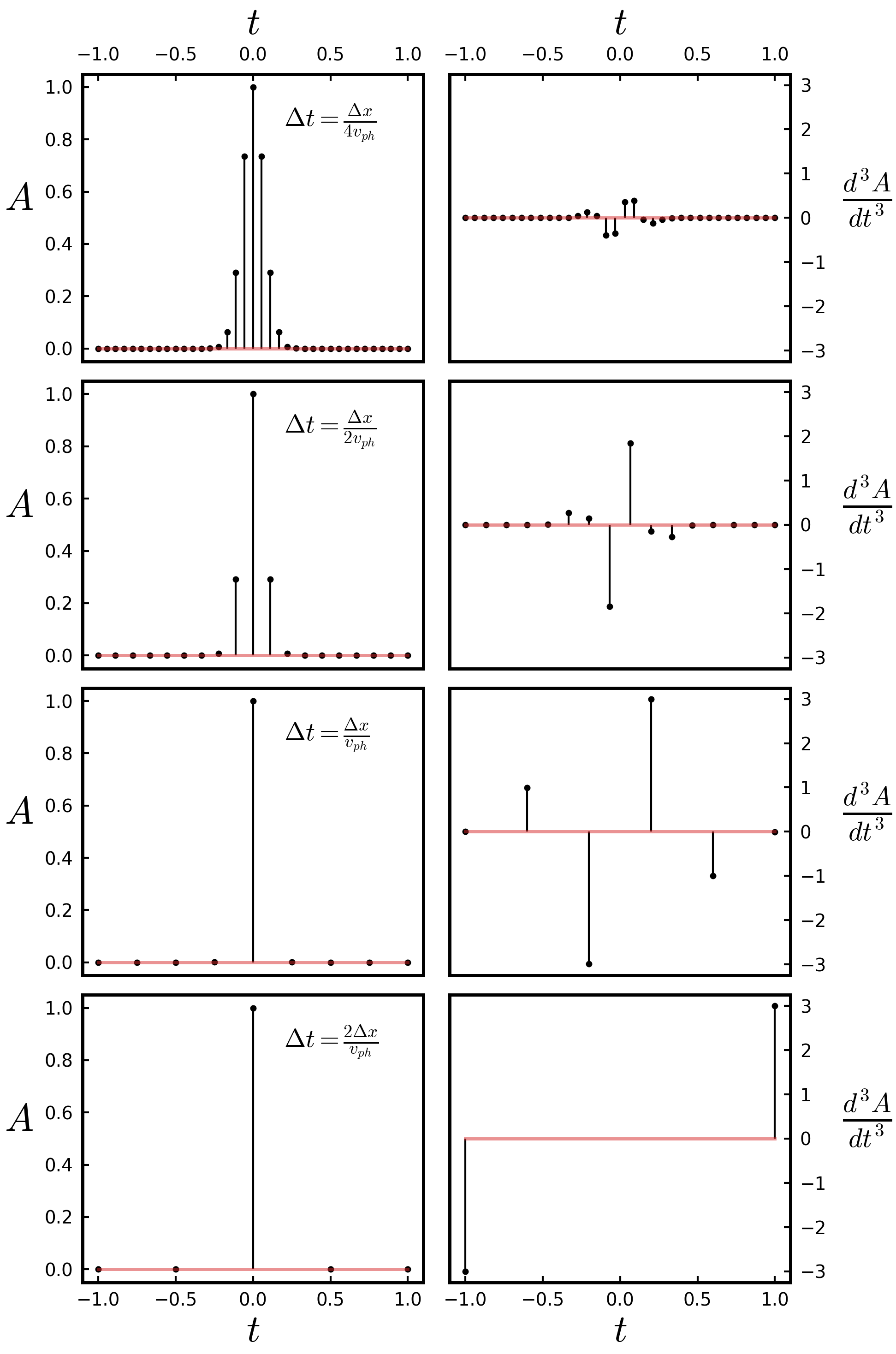}}
\caption{A narrow Gaussian perturbation moving with velocity $v$ is sampled with different temporal sampling intervals at a fix spatial grid point ({\it left} column). Time $t = 0$ corresponds to the time the peak signal crosses that sampling location. The spatial $\sigma$ of the perturbation is 16 km (the signal is spatially sampled at the Nyquist limit of the MURaM spatial frequency) and the velocity of the perturbation is 8 km/s (photospheric sound speed). When the perturbation is over-sampled (first two rows), application of $3$-difference yields low amplitude at all times ({\it right} column). For under-sampled perturbation ({\it bottom} row), the $3$-difference operation fails to produce a usable $n$-difference signal (see text). When the perturbation is critically sampled ({\it third} row), satisfying the relationship $\Delta t = \Delta x/v_{ph}$, the $3$-difference operation produces the strongest response.} 
\label{fig:figure2}
\end{figure}

These behaviors are illustrated by Figure \ref{fig:figure2} which plots the result of applying a 3-time-step difference filter (first-order forward-difference third-derivative [-1, 3, -3, 1] stencil) to a narrow idealized Gaussian perturbation propagating past a fixed position with speed $v$.  When the signal is well resolved ($\Delta t = \Delta x/4v_{ph}$), the 3-difference timeseries has small amplitude, and when the signal is critically-sampled ($\Delta t = \Delta x/v_{ph}$), it has maximum amplitude. When the the signal is under resolved the characteristic temporal-difference waveform is lost.  In the case illustrated, the maximum amplitude of the under-sampled difference is the same as that for critical sampling (compare the third and fourth rows in Figure~\ref{fig:figure2}), but this is a special case.  For clarity, the signal chosen for the illustration in Figure~\ref{fig:figure2} is perfectly aligned with the sampling interval.  The temporal intervals were chosen so that the peak of the propagating Gaussian is precisely sampled at the fixed spatial location of observation.  Typically that is not the case, and both the waveform shape is lost and the maximum amplitude of the difference drops when the sampling rate is sub-critical.  

\begin{figure}[t!]
\centerline{\includegraphics[width=\linewidth]{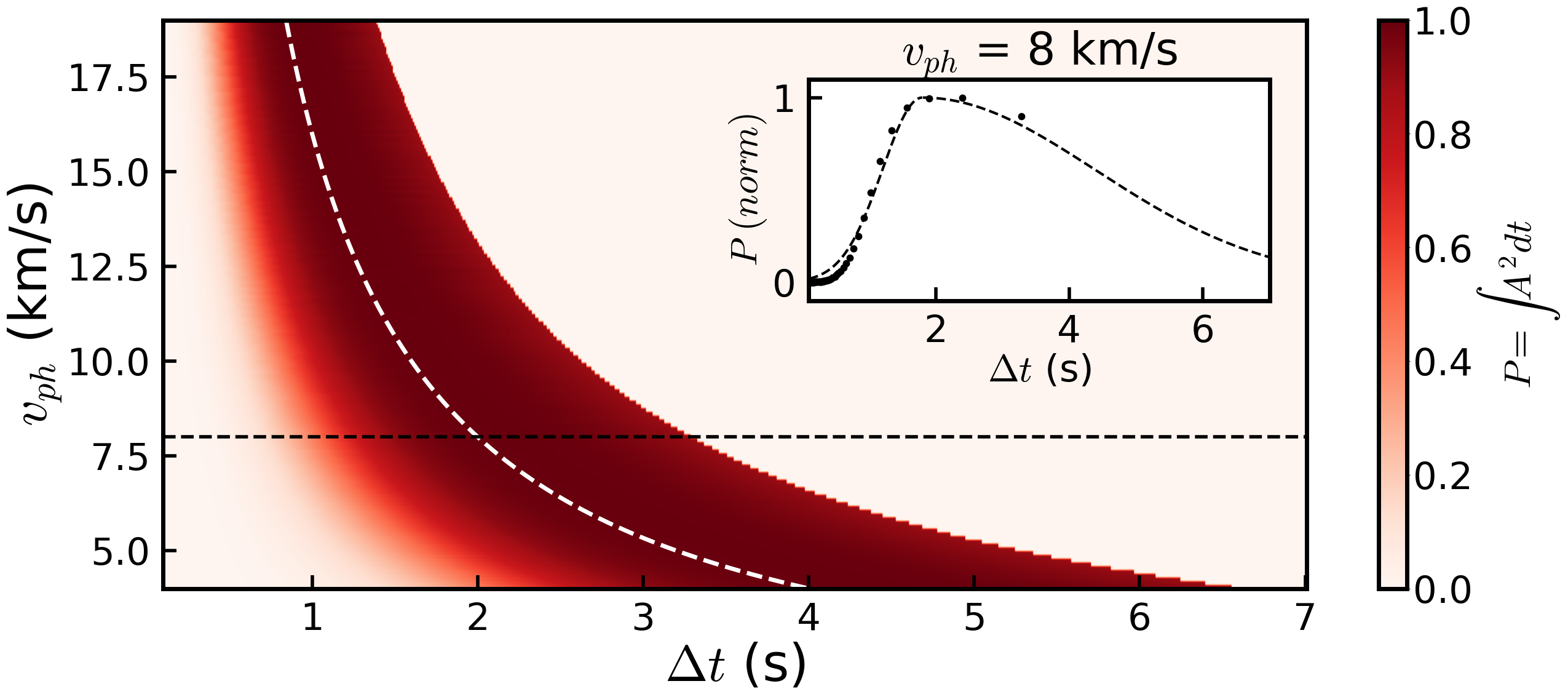}}
\caption{Power (integrated amplitude squared) of the $3$-differenced signal amplitude for varying signal velocity and temporal sampling period in a MURaM grid ($16$ km spatial pixel). The \textit{dashed white} curve represents the criterion $\Delta t = \Delta x/v_{ph}$ where optimal sampling is achieved and the amplitude of the difference signal is maximum. The \textit{dashed horizontal black } line indicates the approximate sound speed at the surface of MURaM photosphere, and intersects the \textit{dashed white} curve at the optimal $\Delta t$ sampling of 2 seconds for a photospheric source.  The phase speed of wavefronts generated by deeper sources are higher and therefore must be sampled at a higher cadence for optimal sampling. \textit{Inset:} Power of the $3$-difference signal amplitude vs temporal sampling period for $v=$ 8 km/s. For clarity, the relation is arbitrarily fit by a third order polynomial (\textit{dashed} line).} 
\label{fig:figure3}
\end{figure}

Thus, the temporal difference filter we propose yields the maximum response when the signal is critically sampled, and the optimal spatial $\Delta x$ and temporal $\Delta t$ samplings are coupled to the phase velocity $v_{ph}$ of the perturbation,  $\Delta t = \Delta x/v_{ph}$.  Since the SNR achieved   
depends on the sampling rate relative to both the signal and noise timescales, the most effective filter over-samples the noise (in our case the granulation) and critically samples the signal (in our case the propagating wavefront). Any perturbation that has $v > v_{ph}$ is sparsely sampled (sub-critical sampling) and yields low amplitude temporal difference on application of the difference filter. Perturbations with $v < v_{ph}$ are over-sampled and also produce low difference amplitudes, and the difference filter can be tuned to isolate propagating disturbances with particular phase speed. The optimal three-difference sampling rate for perturbations of differing phase-speeds are plotted in Figure~\ref{fig:figure3} for fixed 16 km spatial sampling. The difference-signal peaks when the sampling period is critical, and falls away from that value for constant perturbation speed. The observational cadence needed for given spatial resolution can thus be tuned to extract wavefronts of particular phase speed. Note that, contours of constant difference power are smooth for the Figure~\ref{fig:figure3}, but in reality are discrete because the spatial grid spacing is fixed and the signal is aligned with the sampling interval on the grid as the sampling period changes continuously.

\subsection{Filter summary} \label{sec:summary}

In summary, we have improved the wavefront filter previously employed~\citep{2021ApJ...915...36B} to mitigate the drawbacks associated with convolutional operators. This was done by applying the neural network motivated temporal-difference kernel directly to the image time series.  This updated filter has the advantage that it can be tuned to the phase speed of the propagating wavefront, and thus allows observational differentiation of excitation events occurring at differing depths. Explicitly, the filter we employ consists of the following:
\begin{enumerate}
\item{Take three successive differences of time series in time.  In our case the cadence of the simulated MURaM image time series is 2 seconds and the spatial sampling is 16 km, and we difference consecutive images, but \it{a robust observational strategy would be to take an image timeseries with highest spatial resolution and as rapid temporal cadence as possible, subject to the signal-to-noise requirements, and then, in post-processing, to sample that time series with a cadence that critically samples the phase speed of the targeted perturbation}.}
\item{Apply a spatiotemporal Fourier filter to the difference-image time series to remove perturbations with phase speeds greater than 13 km/s. This step is helpful in removing much of the $p$-mode power that can overwhelm the low amplitude signal, but it can also introduce spurious patterns with phase speeds of 13km/s and occasional single-pixel artifacts.  This step is redundant and can be skipped if the interest is only in the highest amplitude wavefronts.  These are readily visible in the difference images without application of the 13 km/s filter (see video files included in Supplementary Materials). In practice, we compare non-Fourier-filtered and Fourier-filtered image time series, using the filtered images for initial discovery of weak signals and the non-filtered images for verification and analysis.} 
\item{Clip large values in the filtered image time series by restricting the residual map to values between $-0.1$ and $0.1$ km/s.  This step saturates any extreme values introduced or amplified by the temporal difference.  Most of these are associated with the evolution and movement of granular/intergranular boundaries.}
\end{enumerate}

\begin{figure*}
\centerline{\includegraphics[width=\textwidth]{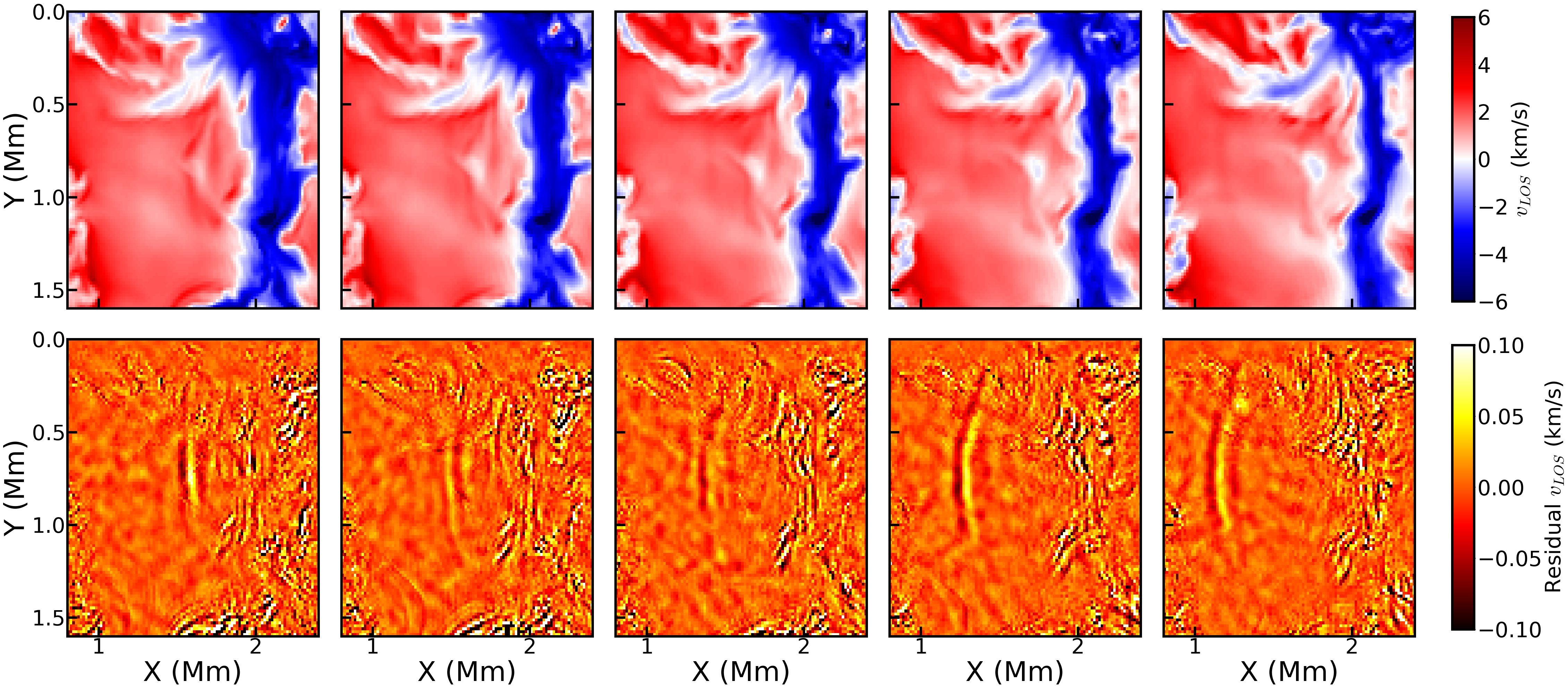}}
\caption{Snapshot of surface line-of-sight velocity ({\it top}) and the corresponding 3-difference temporal filter images ({\it bottom}) illustrating the response of the atmosphere to a strong isolated impulsive source produced by the simulated convections.  Images span 48 s (simulation time stamps ($t = 1588$ s, $1600$ s, $1612$ s, $1624$ s and $1636$ s), and show a small subdomain of run O16b from ~\cite{2014ApJ...789..132R} with non-grey radiative transfer and a domain extended an additional 1.024 Mm upwards into the chromosphere. The timestamps are consistent with those in the video included in the Supplementary Materials.
\label{fig:figure4}}
\end{figure*}

We have applied 3-difference temporal filter described above to reanalyze the same Doppler-image time series used in our previous study~\citep{2021ApJ...915...36B}. We have found that many of the source sites identified are common to the two techniques, though some false positives introduced by the previous convolutional filter have been avoided with the new filter.  Moreover, the general properties of the source locations are common to the two techniques.  Acoustic sources are frequently found in and near intergranular lanes, particularly at locations of complicated mixed flow structure or sudden local downflow enhancement, and 
as previously, multiple sources often occur in close proximity.  However, with the new filtering technique, the strongest and most-visible sources tend to be shallower.  This is because direct temporal differencing favors those perturbations with the largest amplitudes in the photosphere. The statistics of the acoustic sources, their depth distribution and the physical mechanisms which underlying them, are thus sensitive to the particular filter applied, and fully understanding those systemics and careful assessment of selection effects are required before using these techniques to uncover the statistical properties of the sources themselves. We leave that for future work. 

\section{Observational Implications} \label{sec:obs}

In the solar photosphere, an important signature of an impulsive acoustic source at depth is a horizontally propagating wavefront with a time-dependent phase speed that depends on both the source depth and sound-speed profile through which the impulse response has propagated.  Since the photospheric signal is the signature of a three dimensional front propagating through a two-dimensional surface, the deeper the source, the more rapid the initial phase speed as the front pierces the photosphere.  Subsequently, as the wave expands, the front speed asymptotes to the sound speed at the source depth.  The observed photospheric wavefront shape and propagation speed thus captures information about the source depth, the mean atmospheric stratification, and any inhomogeneities in the atmosphere through which the wavefront has propagated.

\subsection{Wavefront propagation}

Application of the the temporal 3-difference filter developed in Section~\ref{sec:Improvement} to a MURaM photospheric Doppler image time-series
reveals that, in the simulation, acoustic sources are frequently found in and near intergranular lanes. {Figure \ref{fig:figure4}} displays the temporal evolution of the photospheric Doppler velocity before ({\it top} row) and after ({\it bottom} row)  application of the difference filter to a region surrounding a strong isolated source. Evident in the time series is an acoustic wavefront, with a width of approximately 6 pixels, propagating across the adjacent granule.

\begin{figure}[t!]
\centerline{\includegraphics[scale=0.27]{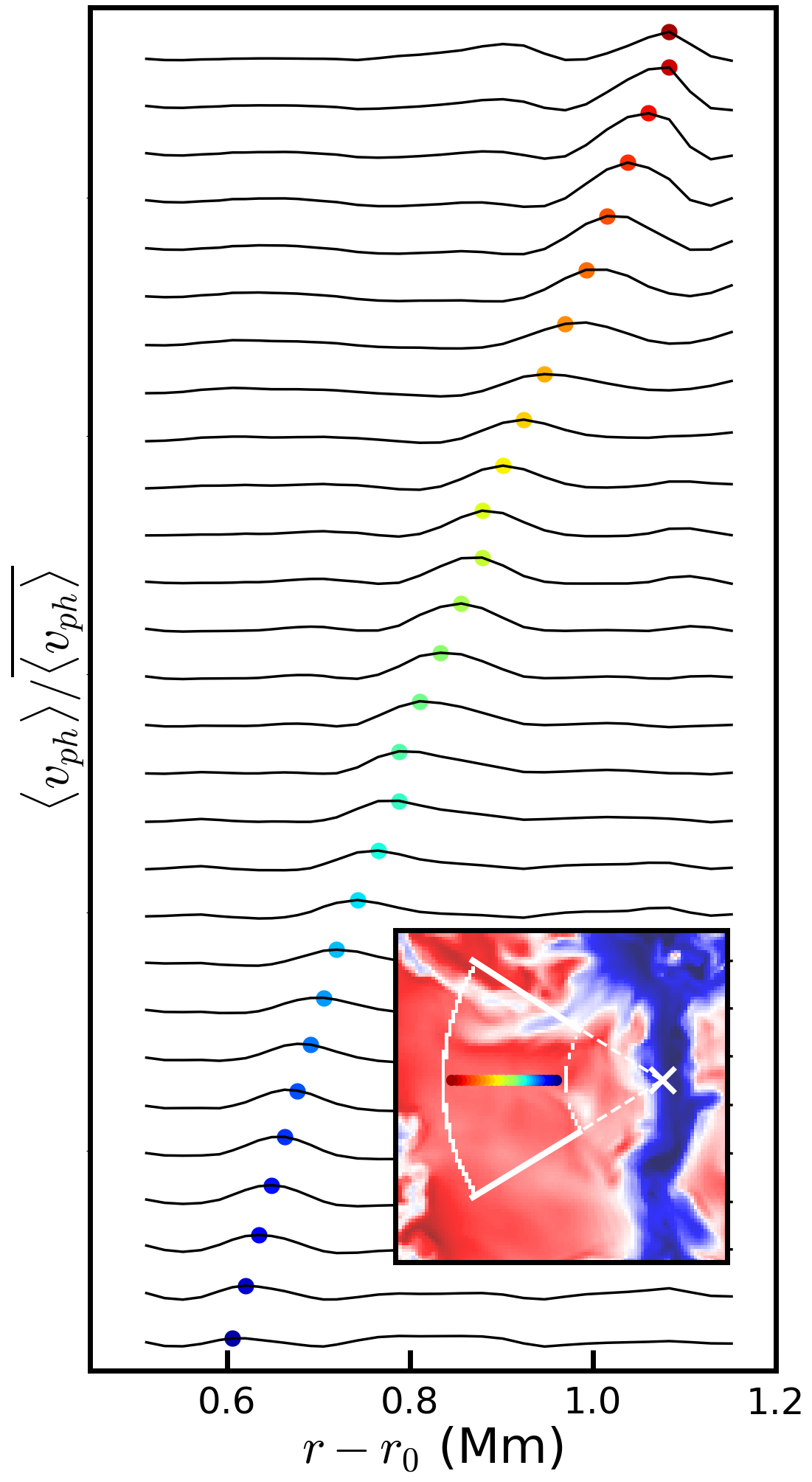}}
\caption{Azimuthally averaged photospheric line-of-sight velocity (normalized by its mean and offset vertically with time) as a function of radial distance from the source site ($r_0$), illustrating the acoustic response of the atmosphere to a strong, isolated source in the intergranular lane.  The plots are stacked vertically with timestamps starting from $t=1596$ s (lowest plot) and ending at $t=1656$ s (uppermost plot), with $\Delta t = 2$ s between them. {\it Inset:} The region over which the azimuthal average was computed, showing the granule at $t = 1614$ seconds and colored markers corresponding to those in the main figure and indicating their radial distance from the source site. The $\times$-mark in the inset indicates the photospheric position of the source assuming an approximately circular wavefront.  The annular region delineated by the inner and outer boundaries between the radial spokes is the region over which source wavefield seismology is conducted in Section~\ref{sec:SWS}.} 
\label{fig:figure5}
\end{figure}

Once the spatiotemporal location of the source in the photosphere (the site of first photospheric emergence of the wavefront) is identified in the difference-filter image timeseries, the wavefront propagation in the unfiltered line-of-sight velocity images can be followed via an azimuthal average.  Figure \ref{fig:figure5} plots the time evolution of the azimuthally averaged flow over a 71 degree subsection of the propagation region (indicated in the figure inset). The time series shows the undispersed propagation of an acoustic wavefront with a spatial scale of about 100 km. The perturbation is launched near the edge of an intergranular lane and propagates across a large granule.  Advection by the granular flow slows the propagation speed for the first $\sim0.15$ Mm, with the initial velocity of the wavefront approximately $ \sim 8$ km/s.  After passing through the center of the granule, the propagation velocity saturates at  $\sim 9.4$ km/s before the front disappears at the opposite side of the granule. 

It is important to note that, as this example shows, it is necessary to account for the horizontal granular flow when determining the intrinsic propagation speed of the wavefront, which can then be used to estimate the source depth (Section~\ref{sec:depth} below).  For the source depth analysis we present in the next section, we approximate the horizontal granular velocity by spatial and temporal average of the photospheric flow: taking the temporal mean over the duration of the event and the spatial mean over a single direction in propagation region, as indicated by the color bar in the inset of {Figure~\ref{fig:figure5}}.  We note that similar correction for horizontal advection when measuring the wavefront speed in high-resolution observations will be difficult, and this difficulty will have to be addressed before observational application of the inversion techniques we demonstrate in Section~\ref{sec:SWS} is possible.  We have future plans to extend the techniques presented Section~\ref{sec:SWS} to allow simultaneously inversion for the source depth, the atmospheric properties, and three-dimensional flow structure, with iteration for the source depth as part of the solution. 

\subsection{Source depth}\label{sec:depth}

To illustrate the dependence of the wavefront properties on the depth of the source, we examine the response of the horizontal-mean-MURaM atmosphere to short and confined perturbations. Investigation of the evolution of the wavefront amplitude, requires solution of the full three-dimensional wave equation governing pressure perturbations in the stratified atmosphere.  Investigations of the wavefront propagation speed, on the other hand, can employ the less computationally intensive ray-tracing approximation, calculating the integrated travel time from the source to the horizontal mean-tau-equals-one surface along rays between the source and that surface.  Here, we
employ a second-order finite-difference spatial-derivative approximation and a fourth order Runge-Kutta time-
integration scheme to solve the full three-dimensional wave equation. This simplified numerical  approach is sufficient given the very short durations of the simulations. The initial condition is taken to be a three dimensional Gaussian with a fullwidth-at-half maximum of 16 km.  Temporally the source is also take to be Gaussian, with a fullwidth-at-half maximum of 2 seconds centered 10 seconds after simulation starts.

\begin{figure}[t!]
\centerline{\includegraphics[scale=0.15]{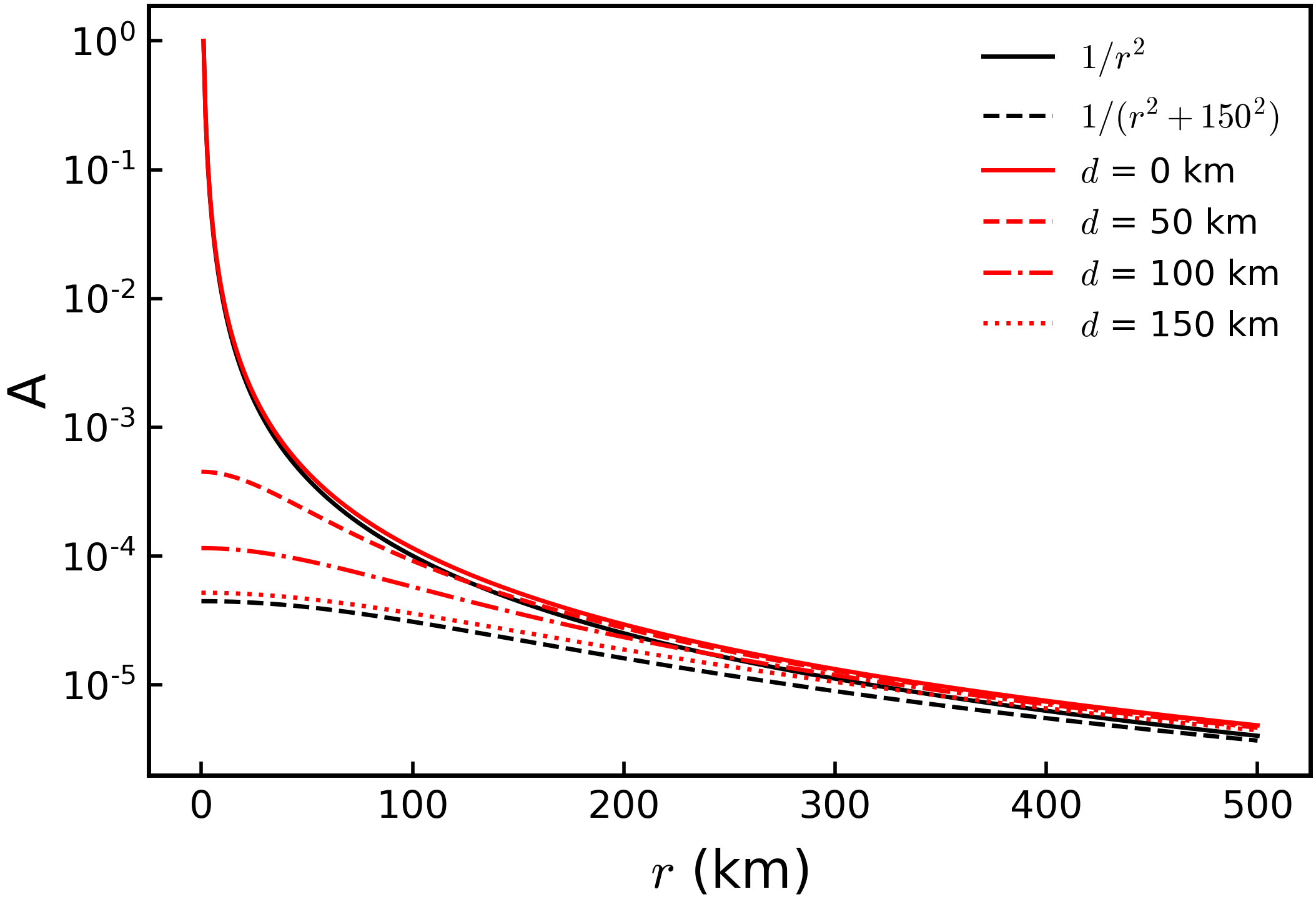}}
\centerline{\includegraphics[scale=0.15]{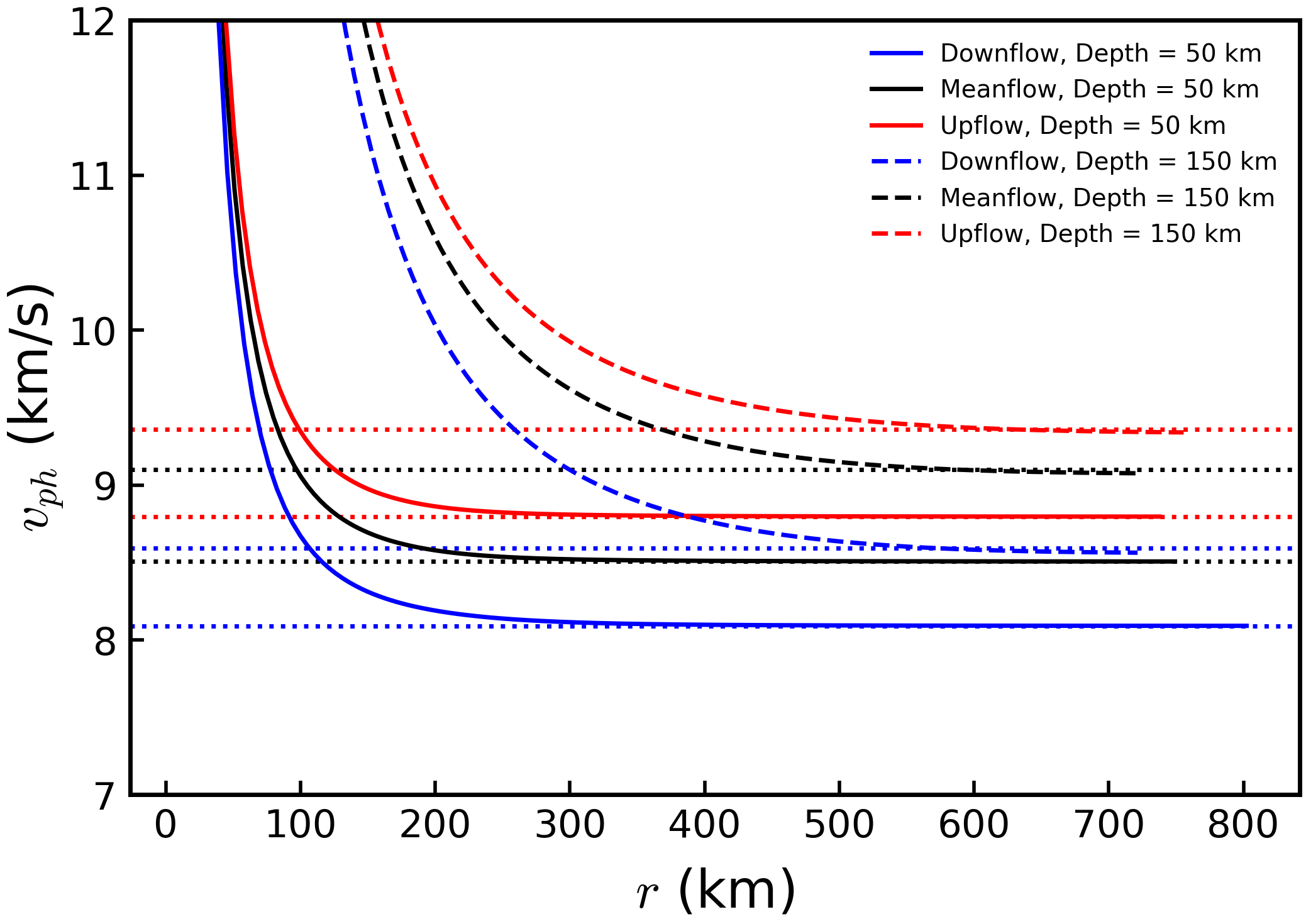}}
\caption{{\it Top}: Change of amplitude $A$ (normalized by the source amplitude) at the photosphere as a function of horizontal distance from the source position, for various source depths. {\it Bottom}: Change in photospheric wavefront phase velocity $v_{ph}$ with distance. As the wavefront passes through the surface, wavefronts from shallower sources undergo faster change in phase velocity and saturate at a lower phase speed (the mean sound speed at the source depth) with distance.}   
\label{fig:figure6}
\end{figure}

The amplitude and speed of the wavefront at mean optical depth unity, as a function of horizontal distance from the projected source location (the location in the photosphere which lies vertically above the source at depth), are plotted in Figure~\ref{fig:figure6}.  Three source depths are illustrated, and the wavefront speed profiles are shown for three different atmospheres:  the horizontal-mean MURaM atmosphere ({\it black}), the horizontal-mean MURaM upflow atmosphere ({\it red}), and the horizontal-mean MURaM downflow atmosphere ({\it blue}).  As expected the amplitude of the wavefront drops as $(r^2+d^2)^{-1}$ (indicated for two source depths, 0 km and 150 km, with the {\it black} curves), with wavefronts from deeper sources showing lower amplitudes in the photosphere as they emerge.  The phase velocity of the wavefront  drops rapidly as it first passes through the photospheric surface and then more slowly with distance after that.  The asymptotic front speed is equal to the mean sound speed at the source depth (indicated in the figure with horizontal {\it dotted} fiducial lines). The deeper the source, the flatter the wavefront as it penetrates the photosphere, and thus the higher the initial phase speed.  The wavefront from deeper sources also takes longer to reach its asymptotic value because a greater distance from the source is required before the ray path can be approximated as horizontal. Thus, if the wavefront is visible for long enough, both the early evolution of the wavefront propagation velocity and its saturated value can be used to determine the depth of the source, and with that and the observed wavefront amplitude the source amplitude at depth.

The wavefront illustrated by {Figure \ref{fig:figure5}}, shows an asymptotic velocity of $\sim 9.4$ km/s after correction for advection by the spatial and temporal mean granular horizontal flow.  In the mean MURaM upflow atmosphere, this corresponds to a source depth of  approximately $160$ km.  Measurement of the wavefront at multiple depths in the MURaM solution, a measurement not possible observationally, suggests that the source depth is $\sim 192$ km.  Both of these measurements are uncertain, and likely in practice determination of the source depth will require simultaneous inversion for it along with the thermodynamic structure of the local atmosphere and the flow velocities with depth. In the next section, we take the first step in demonstrating the feasibility of such inversions, by taking the source depth for this event to be fixed at 192 km and using the observed wavefront speed to invert for the atmospheric properties between that depth and the photosphere.

\subsection{Source wavefield seismology -- proof of concept} \label{sec:SWS}

To illustrated the possibility of using the local acoustic source wavefield in helioseismic inversions, we assume that the acoustic wave motions are adiabatic and that medium varies slowly compared to the wavelength of the sound.  The motion of the wavefront can then be approximated using geometric optics, and we invert the generalized eikonal equation for acoustic waves in a moving inhomogeneous fluid,
\begin{equation} \label{eq_wave}
\vert\nabla\theta\vert=\frac{c_0}{c_s+v_h} \notag
\end{equation}
\noindent
\citep[e.g.,][]{1946ASAJ...18..322B,1953ASAJ...25..950H,1953ASAJ...25..945K,OstashevBook}, for the adiabatic sound speed $c_s$ (reference soundspeed $c_0$) and the flow velocity normal to surfaces of constant phase ($\theta = {\rm constant}$) $v_h$. 
In other words, we find the time-averaged $c_s$ and $v_n$ everywhere within that portion of the domain the wavefront samples, so that the phase velocity of the wavefront $v_{ph}=c_s+v_n$ in the photosphere, as predicted by geometric optics, matches that observed. 

\begin{figure}[t]
\vspace{0.25cm}
\begin{adjustwidth}{-0.5cm}{}
\begin{tabular}{@{}c@{}}
   \includegraphics[width=0.95\linewidth]{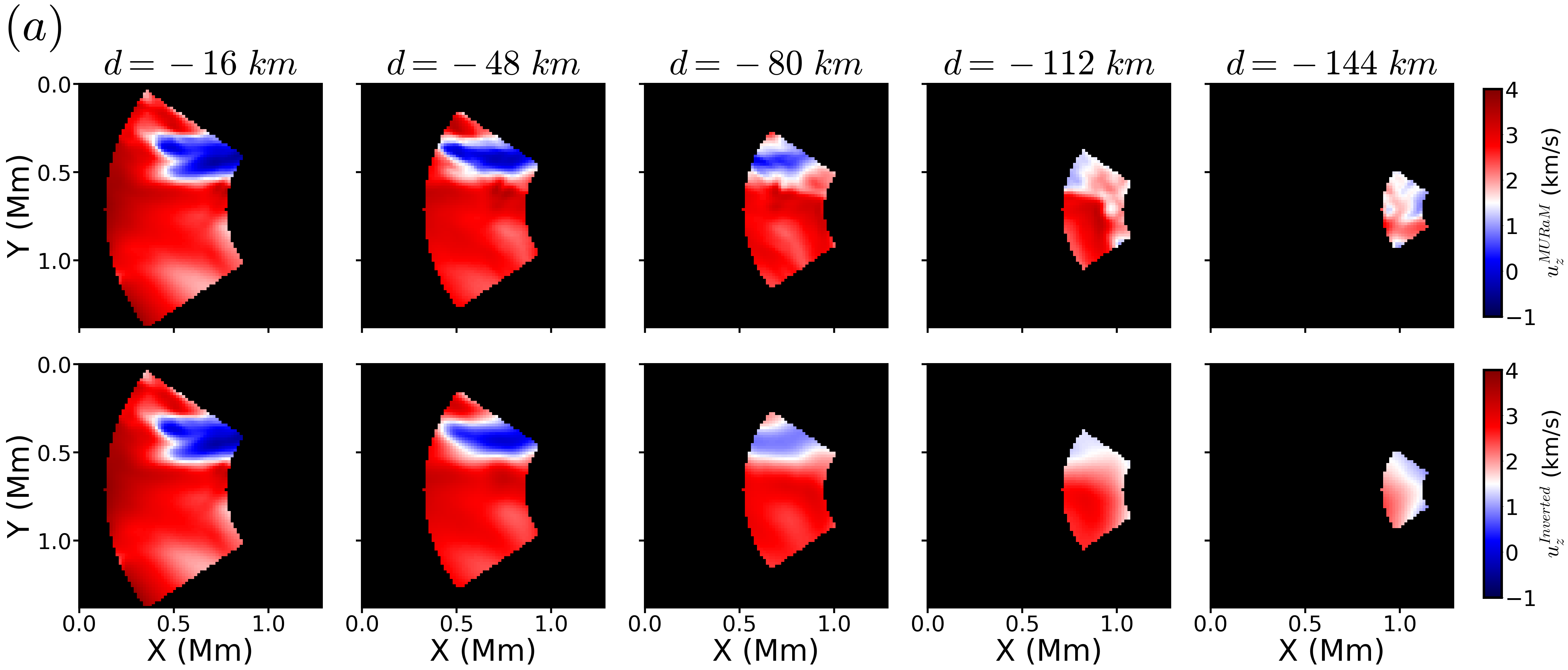}
\end{tabular}
\vspace{0.5cm}
\begin{tabular}{@{}c@{}}
   \includegraphics[width=0.95\linewidth]{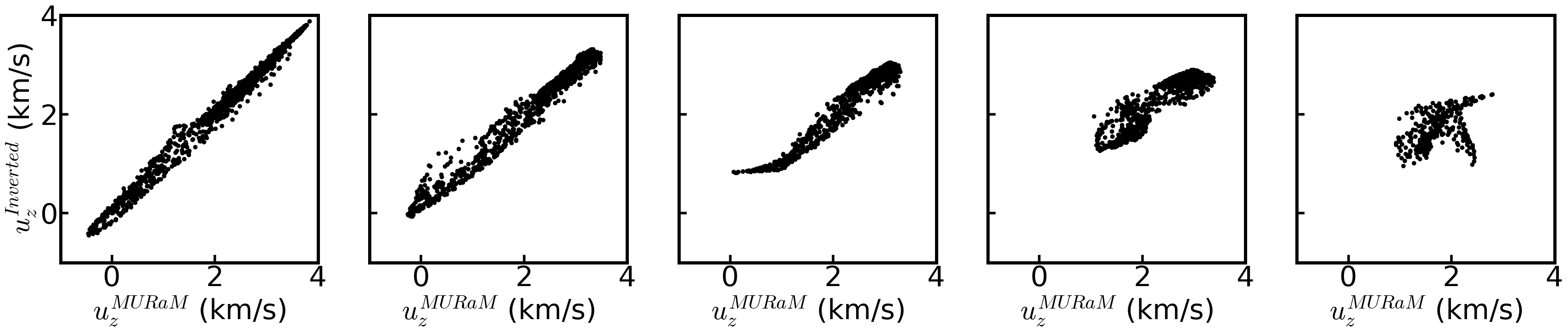}
\end{tabular}

\begin{tabular}{@{}c@{}}
   \includegraphics[width=0.95\linewidth]{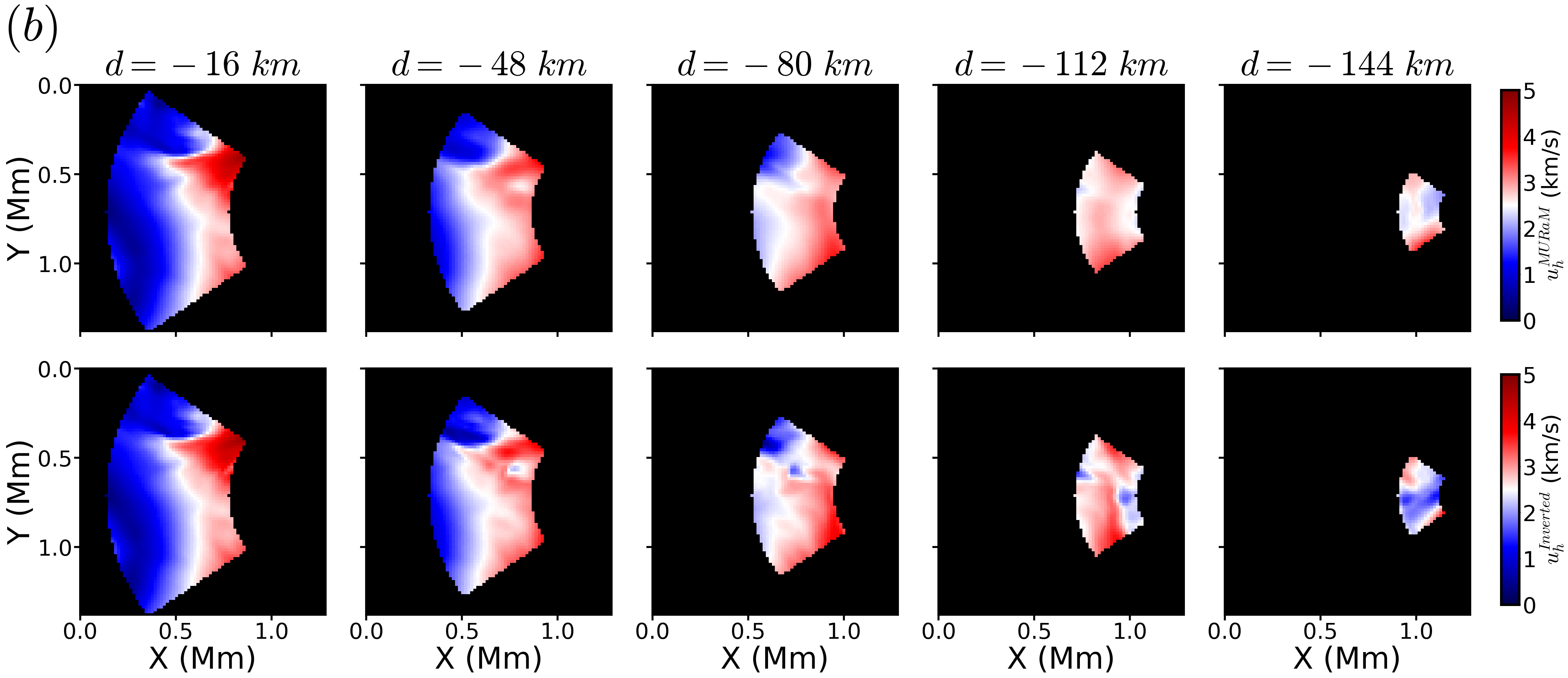}
\end{tabular}
\vspace{0.5cm}
\begin{tabular}{@{}c@{}}
   \includegraphics[width=0.95\linewidth]{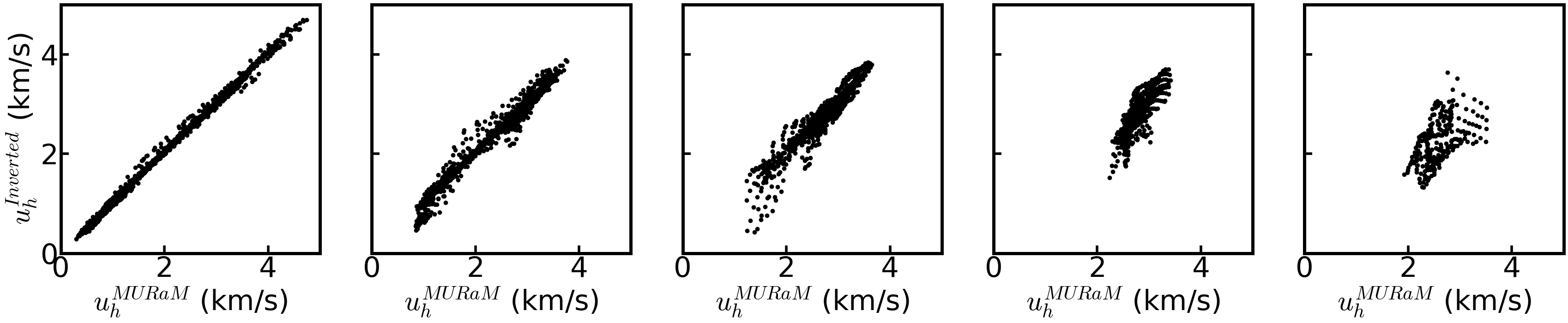}
\end{tabular}

\begin{tabular}{@{}c@{}}
   \includegraphics[width=0.95\linewidth]{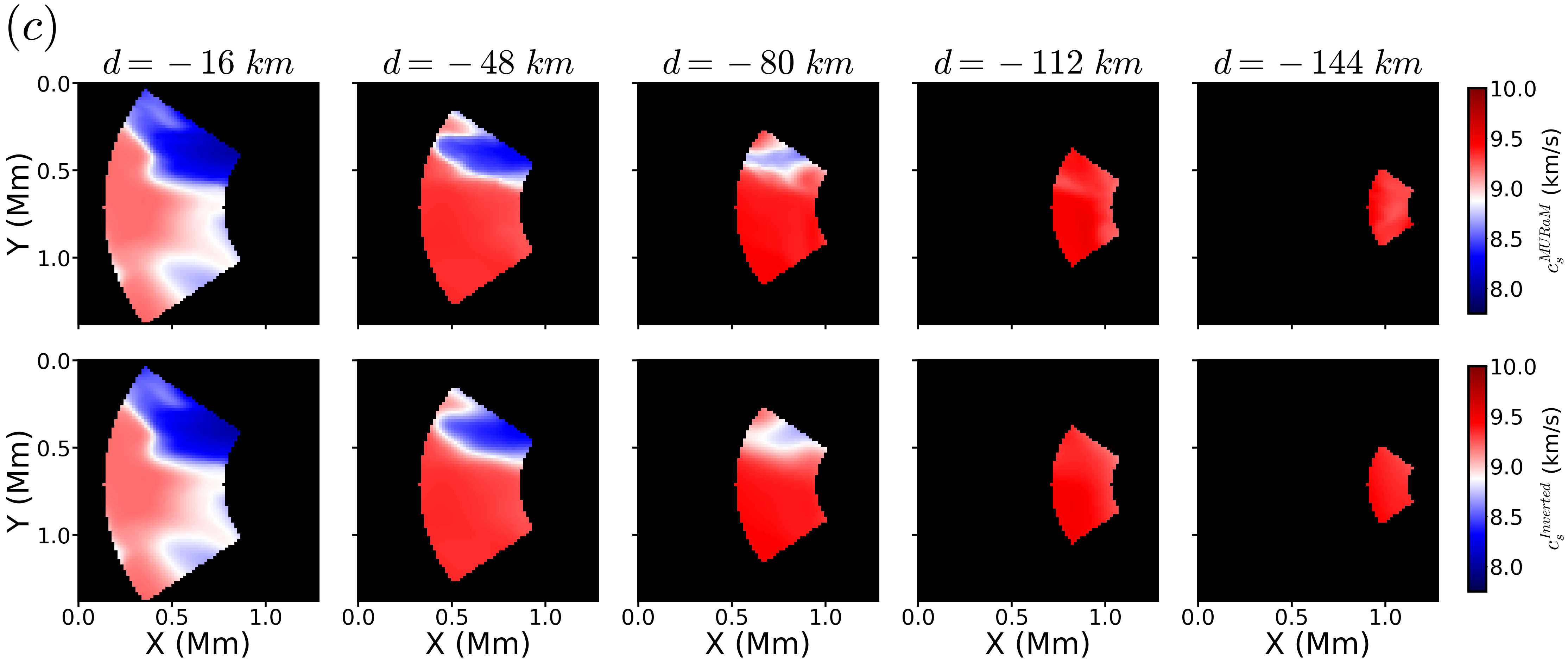}
\end{tabular}

\begin{tabular}{@{}c@{}}
    \includegraphics[width=0.95\linewidth]{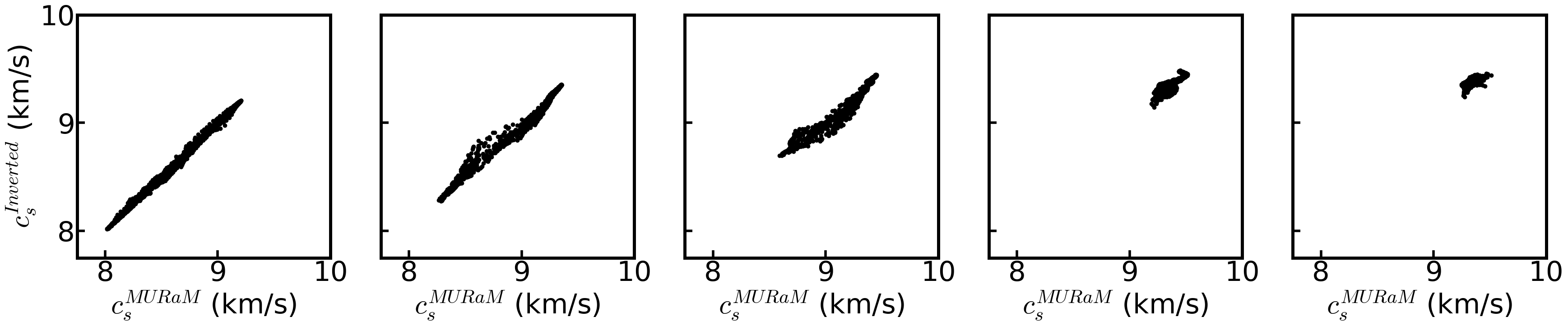}
\end{tabular}
\end{adjustwidth}
\vspace{0.25cm}
\caption{Comparison between the original MURaM atmosphere and the line-of-sight velocity $u_z$ in ($a$), horizontal velocity magnitude $u_h$ in ($b$), and sound speed $c_s$ in ($c$) as deduced from inversions of the source wavefield, at five heights between the source depth (-192 km) and photosphere (0 km).  For each, the simulation values are shown in the {\it top} row, values obtained by inversion are shown in the {\it middle} row, and point-wise correlation between them is plotted in the {\it bottom} row.} 
\label{fig:figure7}

\end{figure}

The inversion proceeds as follows. We assume that the time-averaged vertical velocity and the horizontal velocities follow linear profiles and that the sound speed follows a cubic profile with depth, with each column independent: 
\begin{align} \label{eq_profile} \nonumber
    u_x (x, y, z) = & \ u_x (x, y, z = 0) + m_x(x, y)z \\ \nonumber
    u_y (x, y, z) = & \ u_y (x, y, z = 0) + m_y(x, y)z \\ \nonumber
    u_z (x, y, z) = & \ u_z (x, y, z = 0) + m_z(x, y)z \\ \nonumber
    c_s (x, y, z) = & \ c_s (x, y, z = 0) + m_{c_{s} 1}(x, y)z \\ \nonumber
                    & + m_{c_{s} 2} (x, y) z^2 + m_{c_{s} 3}(x, y) z^3\ . \nonumber 
\end{align}
\noindent
The inversion algorithm adjusts the coefficients of these polynomials, $m_x, m_y, m_z,$ and $m_{c_{s} 1-3}$, iteratively. The region over which the inversion holds is defined by the acoustic ray-paths between the source and the photospheric wavefront position as shown in Figure~\ref{fig:figure5}).  The time-averaged solution is thus valid over an annular three-dimensional wedge decreasing in horizontal extent with height (see Figure~\ref{fig:figure7}).  
 
The inversion is initialized with the location of the source, the time-averaged photospheric line-of-sight and horizontal velocities, and the time-averaged horizontal-mean depth profile of the sound speed. The average velocities in the photospheric wavefront propagation region are held fixed over all iteration cycles of the inversion. The sound speed profile serves only as an initial condition, and the point-wise time-averaged sound speed and flow velocities beneath the photosphere are updated as the inversion progresses. We note that, when conducting similar inversions using solar observations rather than simulations, specification of the source depth and initialization with the time-average horizontal photospheric velocity may prove challenging, and more advanced inversion techniques may be required to obtain a solution.  Here, in our proof-of-concept case, with these quantities well initialized, a simple gradient-descent solver is sufficient to determine the time-average velocity and sound speed at all locations in the inversion region.

Given the initial values of the coefficients and point-wise measurements of the time-averaged velocities and sound speed in the photosphere, a piece-wise linear (between grid boundaries) integration  of the eikonal equation is calculated to obtain the ray path and travel time from source location to photosphere. 
The direction the ray propagates is determined by the spatial variation of time-averaged sound speed, which is accounted for by diffraction at the grid cell boundaries, and by advection by the horizontal and vertical velocities.  The later is determined by the flow velocities in each grid cell and the cell crossing time. The time and location that the rays emerge in the photosphere determines the propagation of the wavefront there, and this can be directly compared to the observed photospheric wavefront.  

Quantitatively, the difference between the two, the loss function, is taken to be the mean-squared-difference, over all image pixels, between the ray-based wavefront image timeseries and a wavefront image timeseries constructed from the MURaM difference images. The wavefront locations in the MURaM difference images are identified as the maxima as a function of radial distance from the source site in 44 radial directions spanning the inversion region.   The $m_x, m_y, m_z,$ and $m_{c_{s} 1-3}$ coefficients are updated to minimize this loss function using gradient descent (Tensorflow's GradientDescentOptimizer function with a learning rate of 0.001) until convergence is achieved.  Convergence is presumed when the loss function falls below $10^{-2}$.

Figure~\ref{fig:figure7} shows the outcome of this procedure: the actual simulated ({\it top} row) and inverted for ({\it middle} row) values of the vertical velocity $u_z$ in ($a$), horizontal velocity magnitude $u_h= \sqrt{u_x^2 + u_y^2}$ in ($b$), and sound speed $c_s$ in ($c$), at five heights in the three-dimensional domain delineated by rays between the source and the observed wavefront propagation region in the photosphere. The correlation between the actual and inverted values at each height are illustrated by the {\it bottom} row of each subplot. The values deduced from the photospheric wavefront inversions are close those of the true time-averaged atmosphere over a significant range of depths. Even with simplifications made in this proof-of-concept test, the correlation between the two remains high to at least 80 km, and while the accuracy of the inversion decreases with depth, it is provides some useful information, particularly for the vertical flow and local sound speed, down to a depth of about 100 km.

\begin{figure}[t!]
\centerline{\includegraphics[width=1.0\linewidth]{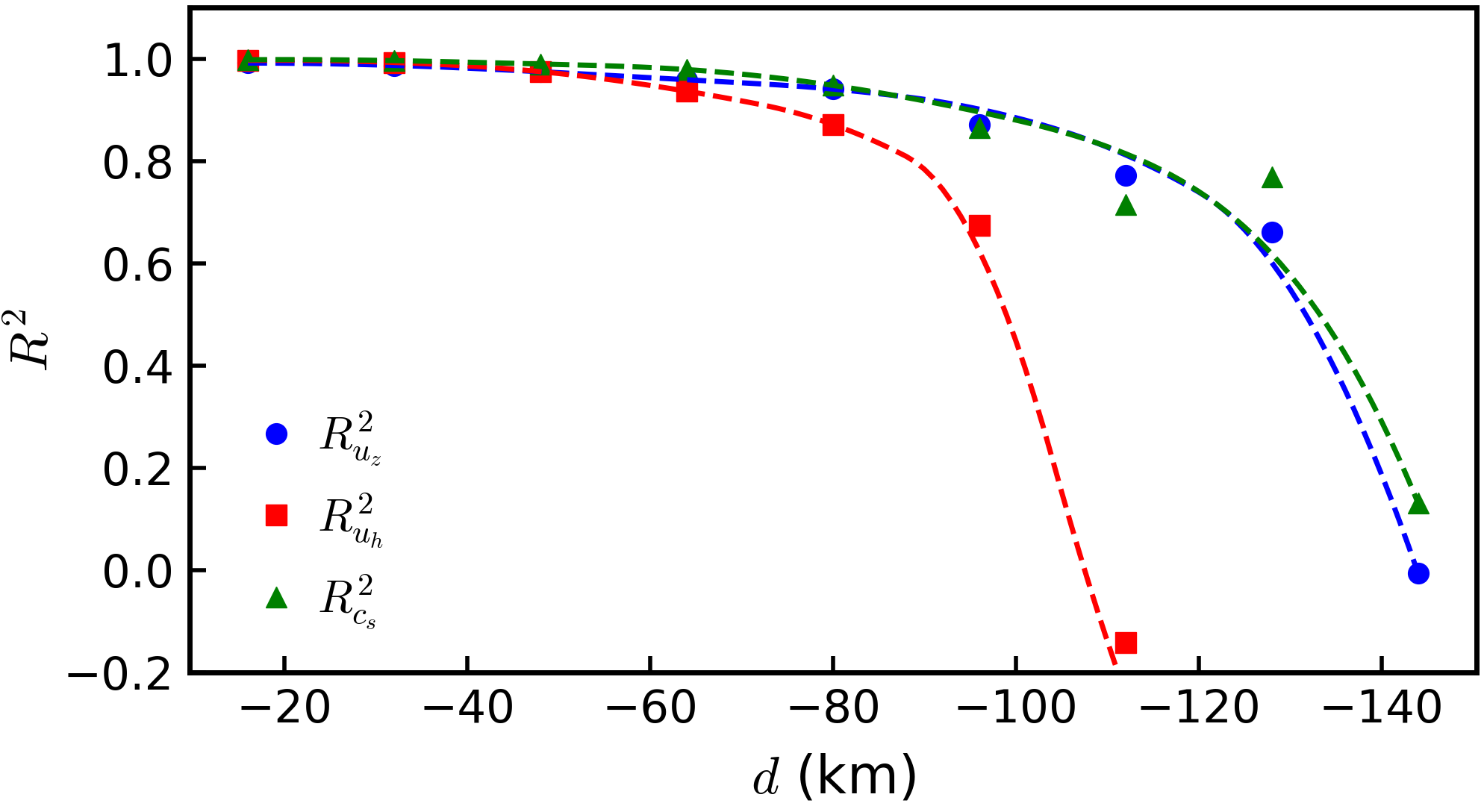}}
\caption{Coefficient of determination ($R^2$) between MURaM atmospheric parameters and inverted parameters are calculated as a measure of strength of the model. The correlation rapidly decreases with depth below $d \approx -80$ km of the photosphere.} 
\label{fig:figure8}
\end{figure}

Figure~\ref{fig:figure8} plots the correlation between the atmospheric properties deduced from the inversion and the MURaM atmosphere as a function of depth.  We note that this reflects both the inversion accuracy and the depth to which the assumed polynomial profiles successfully replicate the MURaM atmosphere below the photosphere. We have found that it is necessary to use a higher order polynomial approximation in the inversion for sound speed in order to achieve an inversion that captures the change in the MURaM simulation in this granular region, while a linear approximation is sufficient for the vertical velocity. This is likely due to the dominant role of  stratification in determining the vertical flow amplitudes. With these polynomial approximations inversion results begin to deviate from the true profiles beyond about 80 - 100 km depth. This is due to the in accuracy of that approximation and the diminishing importance of the atmospheric properties near the source in determining the photospheric wavefront motion.

The inversion appears to more poorly model the horizontal-flow amplitude with depth than it does the sound speed or vertical velocity. The correlation between the actual horizontal flow (computed independently in $x$ and $y$, although here only the flow amplitude is illustrated) and that deduced from the inversion deteriorates rapidly below 80 km. A higher order polynomial in $z$ was attempted but did not yield significant improvement. The assumption of independent vertical column profiles in the inversion may be the underlying cause. Inversion for the horizontal flows likely requires enforcement of horizontal continuity, motivating fully three-dimensional inversions in the future.  

\section{Conclusion} \label{sec:Conclusion}

In this work, we reported on improvements to an image-time-series filter that was previously developed for the detection of locally generated acoustic perturbations~\citep{2021ApJ...915...36B}.  Specifically, we showed how removal of the convolutional kernel from the filter and direct application of image time-series differencing results in increased fault resilience and robustness and improved interpretability. We quantified the relationships between wavefront phase speed and source depth, and demonstrated that, because of that relationship and the nature of the difference filter, 
the filter can be fine tuned, via critical sampling, to identify the acoustic-source depth.  This leads directly to the possibility of using high spatial and temporal resolution image time series to conduct ultra-local helioseismic inversions, and we demonstrated that this is, at least in principle, possible.  The inversion presented here is a first step in the development of these techniques. Application to solar observations will likely require fully three dimensional inversions that recover the source location and horizontal flow in the photosphere simultaneously with the sound speed and flow field at depth.

While employing these methods in studies of the real Sun will undoubtedly pose new challenges, we anticipate that, with the advent of the National Science Foundation's Daniel K. Inouye Solar Telescope (DKIST), similar measurements will be observationally possible.  This promises new ways of investigating the subsurface dynamics and thermodynamics of granulation, small-scale magnetic field generation, pores, and network elements.




\section{Acknowledgments}
The authors thank M. Rempel for generously sharing MURaM simulation results.
This work was partially supported by National Science Foundation grant awards 1616538 and 2206589 and the National Solar Observatory's DKIST Ambassadors program. The National Solar Observatory is a facility of the National Science Foundation operated under Cooperative Support Agreement number AST-1400450.

\bibliography{references}{}
\bibliographystyle{aasjournal}



\end{document}